\documentstyle[aps,twocolumn]{revtex}
\draft
\begin{document}
\author{Y.V.Fyodorov $^{1,2}$, O.A.Chubykalo$^{3}$,
F.M.Izrailev $^{4,5}$ and G.Casati$^{5}$}

\address{$^1$ Fachbereich Physik, Universit\"{a}t-Gesamthochschule  
Essen\\
 Essen 45117, Germany}

\address{$^2$ Petersburg Nuclear Physics Institute,
Gatchina 188350, St.Petersburg District, Russia}

\address{$^3$ Departamento de Fisica de Materiales, Facultad de
Quimica,\\
Universidad del Pais Vasco, San Sebastian, Spain}

\address{$^4$ Budker Institute of Nuclear Physics,
630090 Novosibirsk, Russia}

\address{$^5$ Universita di Milano, sede di Como, via Lucini
3, 22100 Como, Italy}

\title{Wigner Random Banded Matrices with Sparse Structure:\\
Local Spectral Density of States}

\date{\today}
\maketitle

\begin{abstract}
Random banded matrices with linearly increasing diagonal
elements are recently considered as an attractive model for complex
nuclei and atoms. Apart from early papers by Wigner \cite{Wig}
there were no analytical studies on the subject. In this letter
we present analytical and numerical results for local spectral
density of states (LDOS) for more general case of matrices with
a sparsity inside the band. The crossover from the semicircle
form of LDOS to that given by the Breit-Wigner formula is studied
in detail.
\end{abstract}

\narrowtext

Recently there was a growing interest in the statistical
properties of Random Banded Matrices (RBM) (see, e.g. \cite{rev}
and references therein). These $N\times N$ matrices $H_{nm}$ can be
characterized as those containing nonzero elements only within a
wide band of the size $b\ll N$ around the main diagonal. It was
realized that matrices with such a structure are appreciably different 
in their statistical properties from those forming the classical
Gaussian Ensembles studied by many authors \cite{Bohigas}. More
precisely, the RBM of infinite size do not show the famous effect of 
level repulsion that is the direct consequence of the localized
nature of their eigenvectors \cite{selig,Izr,RBM}.

The most studied type of RBM is that with the zero mean value of all
matrix elements and with the variance given by $\overline{H_{nm}
^{2}}=V^{2}a(\mid n-m\mid/b)$, the function $a(r)$ decaying
exponentially (or faster) at $r\gg 1$. Such matrices were shown
to be relevant for understanding some properties of the Quantum
Kicked Rotator, one of the paradigmatic models in the domain of
Quantum Chaos \cite{Izr}. On the other hand, the matrix $H_{nm}$
can be treated both as a tight-binding Hamiltonian of a quantum particle 
in $1d$ system with long range random hoppings and as
 an adequate model for quasi
$1d$ disordered wires \cite{FM}. In a number of publications
properties of such matrices were studied both numerically
and analytically, see references in \cite{rev}. In particular, it
was found that for
matrices of infinite size any eigenvector has typically finite
number (of the order of $l_{\infty}\propto b^{2}$) of essentially
nonzero components. When the matrix is of a large finite size
$N\gg b\gg 1$, its statistical properties were shown to be determined 
by the scaling parameter $b^{2}/N$ \cite{RBM,FM}. As to the
density of states (DOS) it has generally a
form of the semicircular law \cite{SC}:
\begin{equation}\label{semi}
\rho(E)=
\begin{array}{cc}
\frac{2}{\pi R_{0}^2}\sqrt{R_{0}^{2}-E^{2}} & \mid E\mid \le R_{0}\\
0 & \mid E\mid > R_{0} \end{array}
\end{equation}
with $R_{0} \propto b^{1/2}$ being the semicircle radius.

Another type of the RBM -- that with the mean value of diagonal elements
increasing linearly along the main diagonal:
$\overline{H_{nn}}=\alpha n$ --
has attracted recently a lot of research activity
\cite{Wilk,Cas,Fein1,recent,Rob,Flam}. In a tight-binding analogy
mentioned above these matrices
describe a quantum particle in 1d disordered system subject to
a constant electric field. Another interpretation was developed in
\cite{Cas,recent} where matrices of such kind are considered as
Hamiltonians of generic conservative systems with complex behaviour,
like heavy atoms and nuclei.
Indeed, a very detailed study of compound states in the
chaotic $Ce$ atom \cite{Flam,FGG95} has revealed that the
latter
class of the RBM can be used rather efficiently to describe such
a physical system. This particular kind of RBM was
introduced for the first time
by Wigner about forty years ago \cite{Wig}. For this reason
we refer to this ensemble as that formed by Wigner Random Banded
Matrices (WRBM). Quite close ideas were developed recently
\cite{Rob} where the model for a general integrable Hamiltonian
$H_{0}$ perturbed by some generic perturbation
$V_0$ was investigated.
It was argued that the
matrix $(H_{0}+ V_{0})_{nm} $ in the basis of semiclassical
eigenstates of the unperturbed part $ H_{0}$  has the form
of {\it sparse} WRBM.

 As is known, much more informative characteristics compared to the  
global DOS is the so-called {\it local spectral density of states}
(in the literature, LDOS). The latter function, also known in
nuclear physics as {\it strength function}, is defined as
\begin{equation}\label{rho}
\rho(E,n)=\sum_{\nu}\mid \psi_{\nu}(n)\mid ^{2}\delta (E-E_{\nu})
\end{equation}
where $\psi_{\nu}(n)$ is $n$-th component of the eigenvector
$\psi_ {\nu}$ corresponding to the energy level $E_{\nu}$, $n=1,2,...N$.
 After averaging over the disorder the LDOS depends (in the limit
$N\to \infty$) on the parameter $z=E-\alpha n$ only:
$\overline{\rho(E,n)}=\rho_{L}(z)$. As a result, the
global DOS is energy-independent:
$\overline{\rho(E)}=\frac{1}{N}\sum_{n=1}^{N}\rho(E,n)
\rightarrow \frac{1}{\alpha N}$
provided the parameter $\alpha$ is small enough for the sum being
replaced by the integral \cite{Leitner}.

 The quantity of the most physical interest
 is the width $\Gamma$ of the LDOS: it generates
 a new length scale $l_{max}=\Gamma/\alpha$, see details in
\cite{FM,Cas,recent}. If one treats linearly
increasing diagonal elements as eigenvalues of an
unperturbed Hamiltonian the off-diagonal elements play a role of
perturbation and the length $l_{max}$ can be interpreted as an effective
number of unperturbed states coupled by the perturbation
\cite{Cas}. In the limit $l_{max}\ll 1$ standard perturbation
theory can be applied since the perturbation is weak.
We will refer to this regime
as the {\it perturbative} one. In the opposite limit $l_{max}\gg 1$
the problem is essentially non-perturbative. The latter case is
our main concern in the present paper.

 In the case of very large $\alpha$ well inside the perturbative  
regime the form of LDOS is determined mostly by the distribution of  
diagonal elements (e.g. the Gaussian). On the other hand, for  
$\alpha$ in the nonperturbative regime the region of
parameters was indicated in \cite{boris} where the Lorentzian shape
\begin{equation}\label{BW}
\rho_{BW}(E,n)=\frac{\Gamma/2\pi}{(E-\alpha n)^{2}+\Gamma^{2}/4}
\end{equation}
is expected with
$\Gamma\approx 2\pi\rho V^2$\cite{note}. Actually, this form of  
LDOS was obtained by Wigner in his early studies of WRBM \cite{Wig}  
as a result of some  limiting procedure.

The equation Eq.(\ref{BW})
 is commonly accepted approximation for the LDOS in nuclear physics
 known as the Breit-Wigner (BW) formula.  It is considered to be  well in
agreement with experimental data for nuclei \cite{Bohr} and complex  
atoms \cite{Flam}. This fact makes the WRBM to be much more attractive  
as a model for physical systems as compared with the full random
matrices.
However, a broader application of this ensemble is restricted
by the absence of a general theoretical understanding of
their properties. Indeed, the method used in Wigner pioneering
papers \cite{Wig} is rather involved, partially heuristic and seems
to be based upon a specific form of the parameters chosen. In particular,
it is quite unclear how universal are Wigner results (e.g., the
Breit-Wigner
form of the LDOS, Eq.(\ref{BW})) against variations of the parameters 
$\alpha, b$, form of the distribution of the off-diagonal elements,  
and, ultimately, the sparsity \cite{note2}.

Motivated by all this, we have performed an analytical consideration of  
the LDOS
for general WRBM with arbitrary degree of sparsity. For this purpose
we consider random symmetric
matrices $H_{nm}=\alpha n \delta_{nm}+t_{nm}$ , where the probability
distribution of $t_{nm}, n\le m$ is given by:
\begin{equation}
{\cal P}(t_{nm})=(1-p_{nm})\delta(t_{nm})+p_{nm}h_{1}(t_{nm})
\end{equation}
with $h_{1}(t)=\frac{1}{V}h(t/V)$ being a
generic distribution with zero
mean and the variance $V^2$. The probability $p_{nm}$ of being nonzero
for any element is taken to be of the form $p_{nm}=
\frac{M}{b}f(\mid n-m\mid/b)$, where $f(t)$ decays exponentially
or faster at infinity and $\sum_{r=0}^{\infty}\frac{1}{b}f(r/b)=1$.
In such a definition the mean number of nonzero
elements per row is just $2 M$ and the quantity $S=b/M$ is natural to  
term "sparsity".

In order to calculate the LDOS let us follow the method used previously
for sparse matrices without band structure \cite{Sparse}.
Relegating all technical details to a more extended publication,
let us just mention that the mean local DOS is expressed in the form  
of a functional integral, which is finally calculated  by the
saddle-point approximation  exploiting the parameters  $b\gg 1$ and  
$N\gg 1$.
Unfortunately, it is hard to get rigorous
estimate for the domain of
validity of the method for different values of $\alpha, M$ and $b$.
However, for the case of {\it fixed} sparsity one can show that it is
 actually exact
everywhere in the nonperturbative regime of the model.

Performing this straightforward, but lengthy calculation  and
treating the discrete variable $\alpha \frac{|n-m|}{b}$ in the limit  
$b\gg 1$ as a continuous one, we express the LDOS $\rho_L(E-\alpha  
n)$ in terms of a function $G(z_1,\omega_1)$ satisfying the equation
\begin{equation}\label{main1}\begin{array}{c}
\frac{\mu}{M}G(z_{1},\omega_{1})=\omega_{1}^{1/2}\int_{-\infty}^
{\infty} du f(\frac{|u-z_{1}|}{\mu})\\ \times
\int_{0}^{\infty}\frac{d\omega}{\omega^{1/2}}
e^{\frac{i}{2}\omega u-G(\omega,u)}\int_{0}^{\infty} d\tau \tau
h(\tau)J_{1}(\tau\sqrt{\omega \omega_{1}})
\end{array}
\end{equation}
where the notations $\mu=\alpha b/V$ and $z_1=z/V$ are used.
The LDOS is expressed in terms of the
solution $G(z_1,\omega_1)$ as follows:
\begin{equation}\label{main2}
\rho_{L}(z)=\frac{1}{V}\tilde{\rho}(z/V);\quad \tilde{\rho}(z_1)=
\frac{1}{2\pi}Re \int_{0}^{\infty}d\omega e^{i\omega z_1/2-G(z_1,\omega)}
\end{equation}
The pair of equations Eqs.(\ref{main1})-(\ref{main2}) constitute
the main analytical result
of the present paper and give the expression for the
LDOS of sparse WRBM in the closed form.

Let us first consider the case of the
{\it fixed number} of nonzero matrix
elements per row: $M=const$ when $b\rightarrow \infty$.
At given $M$ the form of LDOS
depends on the value of parameter $\mu=\alpha b/V$. When this parameter
is of the order of unity the LDOS form can not be
universal, i.e. it should be dependent
on the particular form of the distribution $h(\tau)$ and on $M$.
 However, the important universality emerges
in the limit $\mu\gg 1$. It appears that in the large domain of
$\omega_1,z_1$
the solution of the equation Eq.(\ref{main1}) can be approximately  
written
as $G(z_1,\omega_1)=g_{0}\omega_1$, where $g_{0}$ is independent of  
$z_1$.
Indeed, substituting this expression in the right-hand side of
Eq.(\ref{main1})
we get:
\begin{equation}
g_{0}=\frac{\pi f(0)M}{2\mu}+O(\omega_1\frac{\mu}{z_1^{2}+\mu^{2}})
\end{equation}
that proves that our approximation is self-consistent as long as
$z_1,\omega_1\ll\mu$. Such a form of $G(z_1,\omega_1)$ immediately  
results
in the BW form of the LDOS, Eq.(3), in the region $|z|/V\ll \mu$,
where the width $\Gamma$ of the LDOS is given by:
\begin{equation}\label{8}
\Gamma=\frac{2\pi f(0)VM}{\mu}\equiv 2\pi
f(0) \frac{M}{b}\frac{V^{2}}{\alpha}
\end{equation}
This allows to give the estimate for the maximal localization length
$l_{max}= \Gamma/\alpha\propto MV^{2}/\alpha^{2}b$.
As was pointed out above,
our {\it nonperturbative} treatment (in particular, the saddle point
evaluation of the functional integral) can be valid only if
$l_{max}\gg 1$.
Together with the condition $\mu\gg 1$ this gives the following
restriction
for the region of existence of the BW regime:
\begin{equation}\label{9}
\frac{1}{b}\sqrt{2\pi f(0)}\lesssim  
\frac{\alpha}{V\sqrt{M}}\lesssim\sqrt{2\pi f(0)/b}
\end{equation}

Another case deserving separate investigation is when the {\it sparsity}
$S=b/M$ is fixed at $b\rightarrow \infty$ rather then $M$ itself.
In particular, $S=1$ case corresponds to the
standard WRBM where we expect
our formulas to be in agreement with those derived by Wigner \cite{Wig}.
 Since $b\gg 1$ is equivalent for this case to
  $M\gg 1$ we can search for the solution
of equation for $G(z_1,\omega_1)$ in terms of an expansion with respect
to $1/M$. One can satisfy oneself that at the leading order in
$1/M$ one can put $G(z_1,\omega_1)=1/2g_{\kappa}(z_1/\sqrt{bS})\omega_1 
\sqrt{bS}$
where the function $g_{\kappa}(x)$ is the solution of the following  
equation:
\begin{equation}\label{wig1}
g_{\kappa}(x)=-\frac{1}{\kappa}\int_{-\infty}^{\infty}f(|u-x|/\kappa)\frac{
du}{iu-g_{\kappa}(u)}\quad;\quad Re g_{\kappa}(u)<0.
\end{equation}
where the control parameter $\kappa=\alpha V^{-1}\sqrt{bS}$ is  
introduced.
 The LDOS is given correspondingly by
\begin{equation}\label{wig2}
\rho_{L}(z)=\frac{1}{\pi}Re \frac{1}{iz-R g_{\kappa}(z/R)}\quad;\quad
R^{2}=b V^2/S \end{equation}
After  introducing the function $r(z)=\frac{1}{\pi}(iz-Rg_{\kappa}(z/R))
^{-1}$
the two equations Eqs.(\ref{wig1}-\ref{wig2})
are identical to those derived by
Wigner \cite{Wig} provided $\alpha=1, S=1$ and $f(t)$ is the step
function:
$f(t)=\theta(|t|-1)$.
It is evident that in the
finite sparsity case the form of LDOS is completely
determined by the only parameter $\kappa\propto \alpha b^{1/2}$. This
fact is in agreement with the mentioned numerical observation
\cite{Leitner}.

When $\kappa\ll 1$ one can immediately find from
Eqs.(\ref{wig1}-\ref{wig2})
 that LDOS
is given by the standard semicircular law Eq.(\ref{semi})
with $R_{0}=(8bV^2/ S)^{1/2}
$ as the radius of the semicircle. When
$\kappa$ increases the BW form of the LDOS (\ref{BW}) emerges
, the width $\Gamma$ and the domain of validity of BW formula
being given by the same expressions Eqs.(\ref{8}-\ref{9}). However,  
now the width $\Gamma$ is actually independent of the parameter $b$  
because of $b/M=const$.
One should stress that in the BW regime the value of $\kappa$
is bound because of the condition Eq.(\ref{9}) which can be  
rewritten as
$\sqrt{2\pi f(0) b} \gtrsim \kappa \gtrsim \sqrt{2\pi f(0)}$. This  
domain is
quite narrow for not large enough values of $b$ taken in
numerical experiments \cite{Wilk,Fein1}.

To show the transition from the semicircle regime to that of
the BW we have performed the numerical study  of
the equations Eqs.(\ref{wig1}-\ref{wig2}) together with the direct
computation of the LDOS from eigenvectors of large matrices of size
$N=1000$ and band size $b=10$ without sparsity: $ S=1$. The  
parameter $V$ was kept constant in a way ensuring the semicircle  
radius $R_0$ to be unity
whereas the parameter $\alpha$ used to change $\kappa$.  The  
function $f(|n-m|/b)$
was taken to be a step-function: $f=0$ for $|n-m|>b$.
For such parameters $b^2 \ll N$ and the finite size corrections are  
small. On the other hand, still there is a region for
$\kappa$ where the BW shape for the LDOS is expected.
The results are presented in Fig.1 where the smooth curves are
solutions of Eqs.(\ref{wig1}-\ref{wig2}) and histograms are obtained by
the average of Eq.(\ref{rho}) over number of matrices and over
those eigenvectors that are not sensitive to the finite value of
$N$. From
this figure the whole transition from the semicircle to the
BW form is seen in detail. Unexpected peculiarity of these
data is that in the critical point (for $\kappa \approx 2$)
there is a local minimum in the center of the LDOS.

Specific question is about the form of tails of the LDOS in the
BW regime \cite{Wig,recent,Flam}. It is rather hard task to
extract the analytical expression for these tails from
Eqs.(\ref{wig1}-\ref{wig2}).
 For this reason
additional numerical check has been done for $\kappa=8$ and $R_0=1$,
see Fig.2. From this figure a
sharp transition from the BW to the exponential form is
clearly seen at $E-\alpha n \approx \alpha b = \kappa R_0 \sqrt{8}
\approx 2.8$ .
The detailed analysis
shows that
it
is impossible to distinguish between the pure exponential dependence
and that found in Ref.\cite{Flam} where the Wigner's
expression for the tails \cite{Wig} was corrected (see also
\cite{recent}).

The results obtained here allow
to understand the nature of the {\it scaling} parameters
governing the statistical properties
of sparse WRBM in different regimes. Indeed, statistical properties of
eigenvectors and eigenvalues are expected to be determined by
the ratio $\beta^{\star}=l/l_{max}$ where $l$ is actual
localization length in the unperturbed basis (see details in \cite{Cas}).
Numerical experiments \cite{Wilk,Cas,Fein1,recent} for the
standard WRBM indeed revealed
that the behaviour of the system is completely determined by the
value of the {\it scaling parameter}
$\lambda=l_{\infty}/l_{max}$ where $l_{\infty}$ is the
localization length $l$ for the limit $\alpha=0$.
This parameter can also be called
the "ergodicity parameter" \cite{Cas} since at $\lambda >> 1$ any
eigenstate is spread uniformly over the scale $l_{max}$ and thus occupies
the maximal possible number of available states in phase space
(in this case $\beta^{\star}(\lambda) \approx 1$).

 In \cite{FMS} it was found
that sparsity does not actually affect the basic fact of
proportionality $l_{\infty}\propto b^{2}$ at $\alpha=0$. However,
the constant $C_M= l_{\infty}/ b^{2}$ is expected to be very small  
at small enough values of $M$.
In contrast, in the {\it fixed sparsity} case $M\propto b\gg 1$ one  
typically
has $C_M\sim 1$. In the latter situation the scaling parameter is
easily shown to be $\lambda\propto (\alpha b)^{2}$ for the BW
regime. Then the condition $\kappa\gg 1$ is equivalent to
$\lambda\gg b$, immediately showing that the BW
regime corresponds to the complete ergodicity: $l_{\infty}\gg
l_{max}$, and
as a consequence to the Wigner-Dyson level statistics (see discussion in 
\cite{recent}). However, if the number $M$ is fixed rather than the ratio
$b/M$, one finds  $\lambda\propto C_M(\alpha b^{3/2})^{2}$ for the  
BW domain restricted by $\lambda \gg C_M b$. For very sparse
matrices the latter condition is compatible with "non-ergodicity"
criteria $\lambda \lesssim 1$ in view of $C_M\ll 1$. Indeed, in the  
paper \cite{Rob} where the extreme case $M=1$ was studied, the
level repulsion was found
to be quite weak and dependent on the combination $\alpha b^{3/2}$
in agreement with the argumentation presented above.

Authors are very grateful to  B.Chirikov, G.Gribakin and I.Guarneri  
for stimulating discussions. YVF, OAC and FMI acknowledge with  
thanks the kind hospitality of the Como
University where this work was initiated.  This work
was supported by SFB 237 "Unordnung
und grosse Fluktuationen", by Grant No RB7000 from the
International Science Foundation and by Grant No 94-2058 from INTAS.

FIGURE CAPTIONS

Fig.1 Dependence of the LDOS on the control parameter $\kappa$.

Fig.2 The tail of the LDOS for $\kappa=8$ and $R_0=1$. The smooth  
solid curve
is the solution of Eqs.(\ref{wig1}-\ref{wig2});
the dotted curve (a) is the Breit-Wigner formula (\ref{BW}); the dotted
straight line (b) is the exponential dependence.


\begin{thebibliography}{99}
\bibitem{note} The expression (\ref{BW}) is valid only as long as
$|E-\alpha n|\ll b$.
In the opposite limit $|E-\alpha n|\gg  b$ the correct expression for  
the LDOS
was derived recently in \cite{Flam}.
\end{thebibliography}
\end{document}